\documentclass[]{aa}
\usepackage{graphicx}
\usepackage{txfonts}
\def\lae{\mathrel{<\kern-1.0em\lower0.9ex\hbox{$\sim$}}}
\def\gae{\mathrel{>\kern-1.0em\lower0.9ex\hbox{$\sim$}}}
\newcommand{\Ly}{Ly$\alpha$}
\begin{document}
   \title{Extended emission around GPS radio sources}

   \author{C. Stanghellini\inst{1} \and C.P. O'Dea\inst{2} \and
           D. Dallacasa\inst{1,4} \and P. Cassaro\inst{6} \and
           S.A. Baum\inst{3} \and R. Fanti\inst{1,5} \and
           C. Fanti\inst{1,5} 
          }

   \offprints{C. Stanghellini}
\institute{
            Istituto di Radioastronomia - INAF, Via Gobetti 101,
	    I-40129 Bologna, Italy 
            \email{cstan@ira.cnr.it}
	    \and
            Department of Physics,
            Rochester Institute of Technology, 54 Lomb Memorial Drive,
	    Rochester, NY 14623 
            \and
            Center for Imaging Science,
            Rochester Institute of Technology, 84 Lomb Memorial Drive,
	    Rochester, NY 14623
            \and
	    Dipartimento di Astronomia, Universit\`a degli Studi, via
	    Ranzani 1, I-40127 Bologna, Italy
	    \and
            Dipartimento di Fisica, Universit\`a degli Studi, via
	    Irnerio 46, I-40126 Bologna, Italy 
            \and
	    Istituto di Radioastronomia - INAF, C.P. 141, I-96017
	    Noto SR, Italy 
}

   \date{Received ............................; 
         accepted 22/06/2005}

\abstract{ 

Extended radio emission detected around a sample of GHz Peaked
Spectrum (GPS) radio sources is discussed.  
Evidence for extended emission which is related to the GPS source 
is found in 6 objects out of 33. 
Three objects are associated with quasars with core-jet pc-scale
morphology, and three are identified with galaxies with symmetric
(CSO) radio morphology. We conclude that the core-jet GPS quasars 
are likely to be beamed objects with a continuous supply of energy 
from the core to the kpc scale. It is also possible that low surface brightness
extended radio emission is present in other GPS quasars but
the emission is below our detection limit due to the high redshifts of the objects.
On the other hand, the CSO/galaxies with extended large scale emission 
may be rejuvenated sources where the extended emission is the relic of 
previous activity. 
In general, the presence of large scale emission associated with GPS 
{\it galaxies\/} is uncommon,  suggesting  that in the context of the 
recurrent activity model, the time scale between subsequent bursts 
is in general longer than the radiative lifetime of the radio 
emission from the earlier activity ($\sim 10^8$ yrs). 
	       }

   \maketitle
%
\section{ Introduction}

The GHz-peaked-spectrum (GPS) radio sources are powerful but compact
($< 1$ kpc) radio sources whose spectra are generally simple and
convex with a peak near 1 GHz. The GPS sources make up $\sim 10\%$
of the bright extragalactic source population at cm
wavelengths. Common characteristics of the bright GPS radio sources
are: high radio luminosity, low fractional polarization, and,
apparently, low variability. 
The optical identifications include both quasars and galaxies.
The galaxies tend to be $L_*$ or brighter and at redshifts $0.1 \lae z
\lae 1$  (O'Dea et al. 1996) while quasars are often found at  very
large redshifts $1 \lae z \lae 4$. 
A review of GPS radio sources has been presented by O'Dea (1998).

Two basic models have been proposed to explain the sub-galactic sizes
of the GPS and the slightly larger CSS (Compact Steep Spectrum) radio
sources. \\
(1) In the {\it frustration} model (e.g., van Breugel et al. 1984; De Young 1993;
Carvalho 1994, 1998) the radio
emitting plasma is confined (for the lifetime of the radio emission)
to a region within the host galaxy by an external medium which is
dense and/or clumpy enough to prevent the expansion of the radio source. \\
(2) The {\it youth} model (Phillips and Mutel 1982; 
Fanti et al. 1995, Readhead et al. 1996,
O'Dea \& Baum 1997; Alexander 2000; Snellen et al. 2000, 2003)
suggests that  GPS
sources with double or triple morphology are small because they are
young and that during the course of their life they will propagate
outwards while decreasing in radio luminosity by about one order of
magnitude for a factor 100 in size. \\
Both the the "youth" and the "frustration" scenarios are motivated
by  the morphological similarity between the small GPS/CSS galaxies
and the extended radio galaxies.

The detection of arc-second scale faint extended emission around
0108+388 (Baum et al. 1990) and other GPS radio sources (Stanghellini
et al. 1990), motivated the suggestion, in the framework of the youth
scenario, that nuclear activity is recurrent in these sources. In this
hypothesis, we see the relic of a previous epoch of activity as
faint diffuse emission surrounding the current young nuclear source.

In this paper we present deep VLA L band images of the 1 Jy complete
sample of bright GPS radio sources (Tab. 1) defined by Stanghellini et
al. (1998). We search for extended (tens/hundreds kpc) radio emission
which could be associated with the GPS source. We discuss the characteristics 
of the positive detections,
the likelihood of their association with the GPS source, and the
implications for models of GPS source growth/evolution. 

H$_0$=100 km sec$^{-1}$ Mpc$^{-1}$, and q$_0$=0.5 have been used
throughout this paper.

\section{ The observations}

The observations presented here were carried out with the VLA in B
configuration in 3 separate sessions on 22 August, 10 September, and 21
September 1998, at 1335MHz and 1665 MHz with a bandwidth of 50 and 25
MHz respectively. All the sources of the 1 Jy complete sample were
observed. 
Each source has been observed for a total time ranging from 5 to 20
minutes. The typical rms noise in the final images is in the range
0.1-0.5 mJy, and the angular resolution is about 5 arcsec.

We also used VLBI images for a comparison with the VLA data when
it was necessary for a better understanding of the source structure.
We used VLBI data from Stanghellini et al. (1997, 2001) and
the Radio Reference Frame Image Database
(RRFID) whose calibrated UV data sets were kindly provided by Alan Fey
(http://rorf.usno.navy.mil/rrfid.shtml). The latter were imaged by us
with appropriate weighting and/or tapering to enhance either the
resolution or the sensitivity to extended emission on the pc scale
(see also Fey et al. 1996, for the original images).

We also examined  the images from the NRAO VLA Sky Survey (NVSS) at
1.4 GHz (Condon et al. 1998) to look for indications of extended
emission on the arcmin scale (the resolution of the images from the
NVSS is about 45 arcsec).

\section{Results}

We found several objects with  diffuse extended emission or a secondary
component close to the dominant GPS radio source. Here we discuss these
sources in detail. In Table 1 and Fig. 1-17 we summarize the results 
of the VLA observations.  In sources where extended emission was seen, 
we superimposed the radio image on an optical image either from 
the POSS2 or, when available, from our Nordic Optical Telescope 
observations (Stanghellini et al. 1993). The  reference for
optical information when not specified is from NED (NASA/IPAC
Extragalactic Database).

\begin{figure}
      \includegraphics[width=8.5cm]{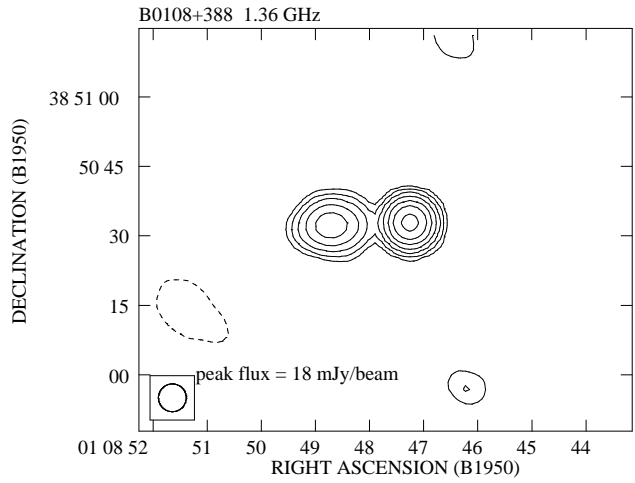}
\caption[]{0108+388 at 1.36 GHz. 0.4 Jy has been subtracted at the core position
to improve the detection of extended emission close to the compact component.}
\label{0108}
\end{figure}

{\it 0108+388}:
the radio galaxy 0108+388 at z= 0.669 is extremely interesting because
it has extended emission at a distance of about 18"
($\sim$80~kpc) from the compact component, in which a  separation
velocity of the micro hot-spots of $\sim0.2c$ has been measured by
Owsianik et al. (1998). Baum et al. 1990 suggested that the extended
radio emission in this object is due to a previous epoch of AGN
activity.  We confirm that there is extended emission only on the
Eastern side of the CSO with a flux density ratio $\sim$25 between the
peak of the detected extended emission and the 3$\sigma$ rms noise
(Fig. \ref{0108}). 

If the extended emission is indeed a relic, there should not be
any side-to-side asymmetry due to relativistic beaming. 
Therefore the one-sidedness of the emission in 0108+388 is puzzling. 
A possible explanation is that
0108+388 was not a classical double but was a head-tail radio source,
of the kind often seen in rich clusters of galaxies (e.g., O'Dea \&
Owen 1985). In this case the one-sided emission would be the relic of
a radio tail. However, there is no evidence that 0108+388 belongs to a
rich cluster of galaxies (Stanghellini et al. 1993).

Another possible explanation is that one of the two lobes has already
vanished below the threshold of detection.  
This may happen if the missing lobe expanded faster in the
intergalactic medium, and experienced stronger adiabatic losses. It is
also possible that the missing lobe faded away faster because of more
efficient radiative losses due to a higher magnetic field on that side
of the radio source. In principle, a difference in the spectral break
due to radiative losses would be expected also if there is a
difference in the travel time between the two lobes, in case the
source axis is not perpendicular to the line of sight. But the double
mas morphology and the lack of relativistic effects suggests the radio
axis is mostly perpendicular to the line of sight and the difference
in travel time is unlikely to produce a ratio $>$25 between the
brightness of the two lobes.

\begin{figure}
      \includegraphics[width=9cm]{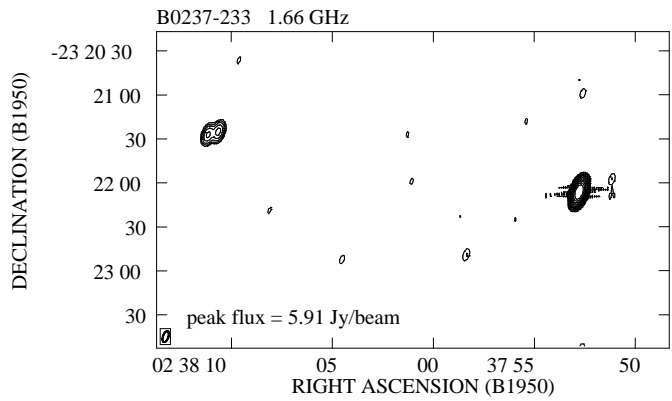}
\caption[]{0237-233 at 1.66 GHz}
\label{0237}
\end{figure}

{\it 0237--233}:
A secondary component (Fig. \ref{0237} ) is seen 5'
($\sim$1.8~Mpc) East of the GPS quasar at $z=2.223$. 
Nothing is visible in the optical POSS2 image at the location 
of the secondary component. However,  the morphology of
the secondary component suggests that this is an unrelated double
radio source. The horizontal stripes visible close to the dominant 
emission are artifacts produced in the cleaning and restore procedure.

\begin{figure}
      \includegraphics[width=8.5cm]{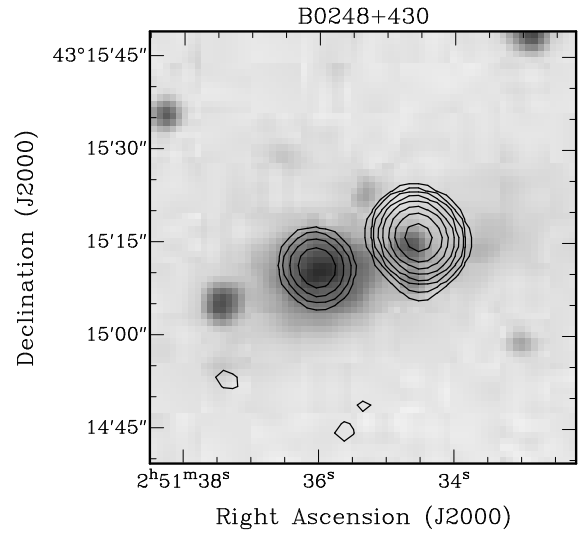}
\caption[]{Radio contours of 0248+430 at 1.36 GHz superimposed to the gray scale POSS2 image }
\label{0248}
\end{figure}

{\it 0248+430}:This quasar at $z=1.310$ exhibits a damped \Ly\ 
absorbing system apparently due to a galaxy at $z=0.3939$.
The absorbing galaxy is clearly seen in the POSS2 plate and is
coincident with a secondary radio component located 20"
($\sim$120~kpc) East of the GPS quasar (Fig. \ref{0248}). 
Thus, it is likely that the secondary radio component is not related to the
GPS quasar, and instead is a foreground object.

\begin{figure}
      \includegraphics[width=8.5cm]{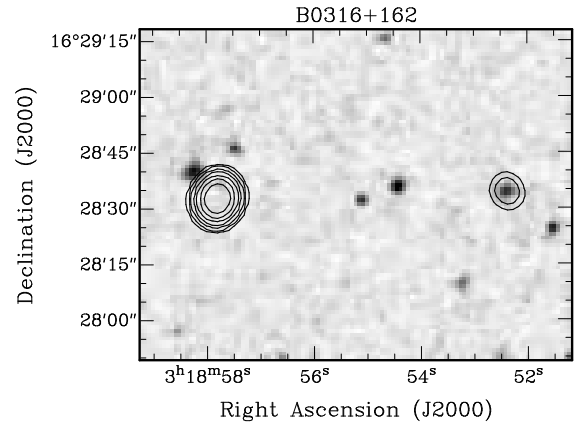}
\caption[]{Radio contours of 0316+161 at 1.36 GHz superimposed to the gray scale POSS2 image}
\label{0316}
\end{figure}

{\it 0316+162}:
A secondary radio component is present at 3' west of the faint GPS
galaxy.  There is an optical object coincident with the secondary component, 
suggesting that it is
an independent radio source (Fig. \ref{0316}). The projected distance
to 0316+162 is about 1 Mpc.

{\it 0500+019}: Stickel et al. (1996) classify the optical host of
this radio source as a quasar, but we consider the identification as a
galaxy at $z=0.583$ made by De Vries et al. (1995) more reliable.
A secondary radio component is present 4' ($\sim$1.1~Mpc) NE of
the GPS radio source. 
No optical counterpart is detected for this secondary component but
its large distance from the GPS radio source gives little support to the
hypothesis that it is related to it.

{\it 0738+313}:This object is a quasar at $z=0.631$ with magnitude
$m_V=16.16$. We find hotspots amidst the lobes on arc second scales
which suggests that the AGN is currently active and is resupplying fresh
electrons to the extended radio components (Fig. \ref{0738}). 
The mas morphology is consistent with a core-jet structure
with the jet showing a highly boosted knot where it changes direction
(helical jet seen in projection?). Detailed radio images of 0738+313
are presented in Siemiginowska et al. (2003) who discuss the nature of
the X-ray emission revealed by Chandra. If we include the
extended radio emission, the total projected source size exceeds 300 kpc.

{\it 0743-006}: The optical identification is a quasar at $z=0.994$
of visual magnitude $m_V$=17.1.
A secondary double component is present 4' ($\sim$1.4~Mpc) SW of
the GPS source. 
There is no optical counterpart but its morphology and distance suggest
it is an unrelated double radio source.

{\it 0941-080}: This GPS source is identified with a galaxy with
a double nucleus at z=0.228. Parsec-scale CSO radio morphology 
has been found by Stanghellini et al. (1993) and Dallacasa et
al. (1998). In our VLA image (Fig. \ref{0941}), a secondary component is present
20" ($\sim$50~kpc) West of the GPS galaxy. 
No optical counterpart is detected for this secondary component. It
has a spectral index of about $-$1.2 between 1.36 and 1.67 GHz. X-ray
emission is detected both at the position of the main and the
secondary components (Siemiginowska, private communication).
Given the proximity to the GPS radio source, the steep spectral index,
and the lack of an optical counterpart we consider 0941-080 a 
candidate for a GPS radio galaxy with extended relic emission on the
arc second scale. 

{\it 1127-145}: Wehrle et al. (1992) have suggested 
that this quasar ($z=1.187$) is a possible compact
double because the mas morphology is dominated by 2 components
separated by about 4 mas. 
We believe the mas morphology of this object is
better classified as a core-jet with a prominent knot
where the jet sharply bends by 45 degrees and continues with additional
bends and wiggles even well beyond the arc second scale. 
The jet can be detected at all the intermediate angular scales from
mas to arcsec (Fig. \ref{1127}) with total extent 25"
($\sim$150~kpc). \\ 
We do not see hot-spots on the arc second scale as we do in 0738+313, 
but the source is at a higher redshift ($z=1.187$) and we possibly
detect only part of the (moderately beamed) jet.
Siemiginowska et al. (2002) report the detection of X-ray emission
associated with this jet. 
The weak radio emission located 3.5' ($\sim$1.2~Mpc) North of
the core is likely an unrelated source as the POSS2 image shows that 
it has an optical counterpart (Fig. \ref{1127b}).

{\it 1245-197}: This radio source is associated with a faint quasar
(m$_V=\sim 21$) at $z=1.275$.
On the 1.36 GHz image (Fig. \ref{1245}) we see a rather compact radio
component 4' ($\sim$1.4~Mpc) West of the dominant GPS radio source. 
The POSS2 red plate superimposed on the radio contours shows a weak
optical feature coincident with this secondary radio emission,
therefore it is likely to be an unrelated radio source.

{\it 1345+125}: This radio source is associated with a galaxy at
$z=0.122$ in which a double optical nucleus is clearly detected
(Stanghellini et al. 1993; Gilmore \& Shaw 1986) . 
Remarkably, the arcsec scale morphology
mimics that seen on the parsec scale, with a southern region bending
to the South-West and a weaker, smaller  northern component
(Fig. \ref{1345}). The GPS source was first detected in X-rays by ASCA
(O'Dea et al. 2000). Newer Observations by Chandra reveal X-ray
emission  coincident with the optical galaxy and extending to the South
partly coincident with the southern radio emission seen by the VLA
(Fig. 3 in Siemiginowska et al. 2004). \\
Reliable information on the spectral index of the lobes/tails between
1.36 and 1.67 is not possible because the rather poor quality of 
the 1.67 GHz data does not allow us to properly image the faint
low surface brightness extended emission. 
This is a good example of a GPS radio galaxy with extended emission on
scales exceeding 1 arcmin ($\sim$100~kpc). However, this radio 
source has rather unusual properties for a CSO. In
addition to the high flux density ratio between the two parsec scale
jets/lobes, it has highly polarized components and superluminal motion
(Lister et al. 2003).  The continuity of the 
morphology from the small to the large scales is consistent with the
hypothesis that there is a continuous supply of energy from the VLBI
core to the extended lobes. 

{\it 1404+286 (OQ208)}: this is a compact radio source associated with
the bright galaxy Mkn668  with m$_r$=14.6 at $z=0.077$ (Stanghellini
et al. 1993). 
This object is one of the closest examples of bright GPS radio
sources. It has a somewhat asymmetric CSO morphology on the pc scale  
(Stanghellini et al. 2001) and there is some indication that it is
expanding (Stanghellini et al. 2003). 
OQ208 is the first radio-loud X-ray Compton-thick AGN ever observed. From ASCA and XMM-Newton observations Guainazzi et al. (2004) estimate a colunm density covering the nuclear emission $> 9 \times 10^{23}$~cm$^{-2}$. The obscuring gas is most likely located within the radio hotspot. A further X-ray absorbing system with $N_H \sim 10^{21}$~cm$^{-2}$ may be associated with gas responsible for free-free absorption of the micro-hotspots.

The detection of a weak (6 mJy) and extended (20",  $\sim$20~kpc)
radio halo has been reported by de Bruyn (1990) from WSRT data at 5
GHz around the compact object. However, we do not see any evidence for 
the extended emission in our VLA images at 1.36 or 1.67 GHz (Fig. \ref{1404}).
Radio emission seen in lower resolution observations may be resolved out with higher 
resolution, though our VLA observations should be sensitive to angular
scales up to a couple of arcminutes.

A secondary component with no optical counterpart is detected 3'
($\sim$180~kpc) NE of the GPS radio galaxy. Its nature is uncertain,
but we suspect it is probably an unrelated object. 

{\it 1518+047}: it is one of the rare quasars (z=1.296) which have a
CSO morphology on the pc scale (Dallacasa et al. 1998). 
The NVSS image suggests the presence of extended emission around this
GPS radio source but the VLA B array image shows an isolated secondary
component about 1' ($\sim$360~kpc) SW of the GPS radio source
(Fig. \ref{1518}). There is no optical counterpart on the POSS2 plate 
and it is unclear if this component is associated with the GPS source.

{\it 2008-068}: this CSO is identified with a galaxy with unknown
redshift.  Using its optical R magnitude of 21.3 and the GPS galaxy
Hubble diagram (O'Dea et al. 1996) we can estimate z$\sim$0.7.
This case is very similar to the previous one. The NVSS indicates
extended emission; however,  the VLA B array image reveals an isolated
secondary compact component  at about 50" ($\sim$250~kpc) to
the SE of the main one, with no optical counterpart
(Fig. \ref{2008}). The spectral index between 1.36 and 1.67 
of this secondary component is slightly inverted so it is
probably an unrelated radio source.

{\it 2128+048}: This CSO is associated with a very red galaxy (magnitude
$m_r=23.3, m_i= 21.85$) by Biretta et al. (1985) with redshift 
0.990 (Stickel et al. 1994).  We detect a slightly resolved radio 
component  4' ($\sim$1.4~Mpc) SE of the GPS radio galaxy with no
optical counterpart (Fig. \ref{2128}).

{\it 2134+004}:
The quasar 2134+004 has a double morphology at mas scale
which has been interpreted in the context of a core-jet structure 
(Stanghellini et al. 2001). 
On the arc second scale it shows
extended emission South of the dominant compact component,  
connected to and likely associated with the GPS source 
(Fig. \ref{2134}),
and an isolated knot at 1.5' ($\sim$540~kpc) from the core, of
uncertain interpretation, and without an optical counterpart on the
POSS2 image.

\vskip 0.5cm

\begin{table*}
\begin{center}
\begin{tabular}{llc|rr|rr}
\hline \hline
\multicolumn{3}{c}{}& \multicolumn{2}{c}{1.36 GHz} &     \multicolumn{2}{c}{1.66 GHz} \\
\hline
name&id & z & peak & extended & peak & extended  \\
& & &mJy & mJy & mJy & mJy  \\
\hline
B0019--000a &G& 0.305 & 2932 & - &  2652 & -  \\
 B0108+388ac&G& 0.669 & 414  & 9.3&  595 & 6.0   \\
B0237--233a &Q& 2.223 & 6282 & (62.0)  & 5950 & (51.5)  \\
 B0248+430a &Q& 1.310 & 1197 & (26.2) &  1281 & (25.1)\\
 B0316+162a &G& (1.57)& 7820 & (25.0) &  7167 & (21.9)  \\
 B0428+205a &G& 0.219 & 3720 & - &  3665 & - \\
 B0457+024a &Q& 2.384 & 2242 & - &   2417 & -\\
 B0457+024c &Q& 2.384 & 2248 & - &    2232 & - \\
 B0500+019a &G& 0.583 & 2144 & (42.1) &   2341 & (30.5) \\
 B0710+439a &G& 0.518 & 1953 & - &     &        \\
 B0738+313a &Q& 0.631 & 2036 & 86 &    2461 & 57 \\
 B0742+103a &Q& 2.624 & 3308 & - &   3795 & -  \\
B0743--006a &Q& 0.994 & 690  & (25.2) &   841 & (17.9)  \\
B0941--080ab&G& 0.228 & 2725 & 31.1 &  2401 & 24.4 \\
 B1031+567a &G& 0.460 & 1791 & - &  1741 & - \\
 B1117+146a &G& 0.362 & 2445 & - &  2153 & -  \\
B1127--145b &Q& 1.187 & 5448 & 96.2 &5371&75.4  \\
B1143--245b &Q& 1.940 & 1557 & -    &1676&- \\
B1245--197b &Q& 1.273 & 5322 &(54.6)& 4708 &(45.4)     \\
 B1323+321a &G& 0.369 & 4855 & - &   4339 & -  \\
 B1345+125ac&G& 0.122 & 5262 & 35.4 &   4734 & 32.1 \\
 B1358+624c &G& 0.431 & 4357 & - &   3860 & - \\
 B1404+286a &G& 0.077 & 792  & (7.3) &   1070 & (5.8)  \\
 B1442+101a &Q& 3.522 & 2432 & - &    2220 & -  \\
 B1518+047a &Q& 1.296 & 3963 & (243) &    3457 & (211)  \\
 B1600+335a &G& (1.10) & 2914 & - &  2834 & -  \\
 B1607+268a &G& 0.473 & 4875 & - & 4387 & - \\
B2008--068a &G& (0.70) & 2576 & (55.4) &   2523 & (58.3)  \\
B2126--158a &Q& 3.270 &  562 & - &  684 & -  \\
 B2128+048a &G& 0.990 & 3970 & (19.2) &  3747 & (18.5)  \\
 B2134+004a &Q& 1.932 & 3412 & 11.0* &  4608 &  7.1* \\
 B2210+016a &G& (0.68)  & 2823 & -  &  2522 & -\\
 B2342+821a &Q& 0.735 & 3797 & -  & 3286 & - \\
 B2352+495a &G& 0.237 & 2532 & -  & 2390 & -  \\
 \hline
\end{tabular}
\end{center}
\caption{ Radio emission data of the sources in the three
epochs of the observations (a= 22 Aug. 1998; b= 10 Sep. 1998; c= 21
Sep. 1998).   
Col. [1], IAU name; col. [2] optical id.; col. [3] redshift
(values in parentheses are photometric redshifts, using the GPS galaxy
Hubble diagram); col. [4]-[7]
flux density of the arc second core and any additional radio emission
at 1.36 and 1.66 GHz
(values in parentheses indicate that in our judgment, the radio
emission likely comes from an unrelated object). Flux density errors
for compact strong components are dominated by the calibration
uncertainties estimated around 3$\%$, while for extended weak
components the flux density errors are dominated by the limits of the
self-cal/imaging process in the presence of a strong source nearby,
and the flux densities given can be considered only rough estimates.
*) it accounts for the region South of the arcsecond core only.}
\end{table*}

\section{Discussion}
We observed all 33 GPS radio sources in the 1 Jy complete
sample defined by Stanghellini et al. (1998). We find 16 objects
showing some extended radio emission closer than either tens of arcsec
or a few arcmin. In general, such additional components are much
weaker than the GPS source. 

Nine sources exhibit radio emission closer than 2 arcmin from the
bright compact GPS source, corresponding to a distance of about
half a Mpc for redshifts above 0.2. 
In {\bf six} of them, 0108+388, 0738+313, 0941-080, 1127-145 and 
1345+125, and 2134+004 the extended emission seems related 
to the GPS source based on the morphological and spectral index information,
possibly due to an earlier episode of activity. 
For the remaining three sources, with additional emission at a
projected distance ranging from about 120 kpc to about 360 kpc, a
physical connection can be excluded in the case of 0248+430 due to the
presence of another optical identification for the confusing source,
and of 2008-068 due to the flat spectral index of the secondary
component. It is unclear how to interpret the additional components in
the field near 1518+047. It seems likely that it is
an unrelated field source, given that it is unresolved by our
observations. 

The weakest secondary component in these 9 fields of 2 arcmin radius, 
is the 9.3 mJy component East of 0108+388.

Based on source counts at 1.4 GHz, there is a $\sim$5\% 
probability of finding a radio source brighter than 10 mJy within 2 arc
minutes from our GPS sources (Prandoni, private communication).  
Thus, we might expect to find about 2 such coincidences by chance.

The radio source 1404+286 has a secondary component at about 3 arcmin distance,
but since it is at a low redshift, the corresponding linear distance is 180 kpc only.
A connection with the main component cannot be completely excluded, but it is unresolved and in this 
case we prefer the intrepretation as an unrelated field source.

Six sources (0237-233, 0316+162, 0500+019, 0743-006, 1245-197 and
2128+048) have possible secondary components at projected distances
exceeding 1 Mpc making the physical connection very unlikely, since
it would imply a total source size two times larger if we consider an
invisible (faded away?) counterpart on the other side of the newly
active GPS nucleus.

In Fig. \ref{0738} we show how 0738+313 (at z=0.631) would appear at
redshifts of z=2 and z=3, keeping constant sensitivity of the
observations (the change of the angular size has not been considered
since at high redshift the angular size changes very smoothly). 
At z=2, 0738+313 would appear as a strong radio source with a
secondary weak component 30" in the north direction, and another
barely visible on the other side of the strong component. At z=3 there
would be no clear evidence of additional radio emission, just a hint
of the northern hotspot emission which could be easily interpreted as
a peak in the noise. 
Due to the redshift distribution of the 14 GPS quasars of
our sample, the extended emission revealed around 0738+313 (which is
the closest quasar of our sample) would be impossible to detect in 
the vast majority of them (Fig. \ref{fluz}).
This indicates that the extended, low surface-brightness lobes of
a classical double may be below the detection limits of our
observations, and only the brightest components (the GPS core and one or two
isolated components, possibly the hot-spots or knots in a jet) are
visible in our images. Deeper observations are needed to search for
fainter emission connecting the GPS source and the secondary components. 

In Fig. \ref{fluz} we plot the flux density we have detected (and
upper limits) from the extended emission (assuming the extended sources 
cover an area of ten times the beam areas) as a function of redshift. 
In the diagram, the regions usually populated by FRI and
FRII-type radio sources (e.g., Ledlow \& Owen 1996) are also marked.
This illustrative example shows that we cannot rule out the
possibility that the GPS sources contain extended FRI and even
FRII emission that remains undetected because our sensitivity rapidly
decreases with redshift. 

In this somewhat heterogeneous picture, three galaxies (out the 19 galaxies in our
sample) appear to have extended emission:  
\begin{itemize}
\item{} 1345+125 where there is continuity between the small
  and the large scale with a (twin) jet progressively diffusing into a
  low surface brightness tail.

\item{} 0108+388 in which the youth of the small scale radio emission
  has been confirmed by the measure of the advance speed of the
  micro-hotspots (Owsianik et al. 1998), and there is resolved, steep
  spectrum old emission on the large scale. This is a likely case of
  recurrent activity as already proposed by Baum et al. (1990).

\item{} 0941-080 where X-rays have been detected from the
  additional component, complicating the interpretation. 
\end{itemize}

On the other hand, the three quasars have a core-jet morphology 
on the small scale while on the large scale we see either an FR-II type radio
source (0738+313) or the continuation of the small scale jet
(1127-145) or a more complex structure (2134+004). 

Why do we see different relationships between the extended emission 
and the compact 
GPS source? It has been clear for a long time that the selection of
sources on the basis of a peaked spectrum produces a somewhat
heterogeneous sample of objects (O'Dea, Baum, Stanghellini 1991). So 
we should not be surprised to find different relationships among the
sources discussed here.

\subsection{\bf Extended emission in GPS galaxies}
 
Among the 19 galaxies of the complete sample, the only clear case
where the recurrent hypothesis can be applied is that of 0108+388,
whose youth has been confirmed by the kinematic age measurement
(Owsianik et al. 1998). A similar case could be 0941--080, but the 
kinematic age is not yet available and we know little about the
properties of the additional component. \\
Finally, the source 1345+125 shows evidence for continuity between the
small and large scale emission, which, if confirmed, would suggest that the
extended emission is not  the relic of past activity. \\ 
The total fraction of GPS {\it galaxies\/} with extended emission is therefore
between 5--16\% (1--3 out of 19). 
We can therefore conclude that in general, the presence of large scale
emission associated with GPS galaxies is quite rare, and
all but one of the cases found so far still are subject to the possibility 
that they could arise
from chance projection effects rather than a physical connection. 
In the context of the recurrent activity model, all this suggests that 
the time scale between subsequent bursts is in general longer than the
radiative lifetime of the radio emission from the earlier activity
($\sim 10^8$ yrs). 

In several sources we may be witnessing ongoing (or at least recent)
transport of radio plasma from the core to the  extended structure. 
In the case of the GPS quasar 0738+313, with a core-jet mas morphology,
we see the hotspots on arcsecond scales which suggest that there is
a continuous supply of fresh electrons from the nucleus. 
In the quasar 1127-145 which also has core-jet mas morphology, 
we see radio emission from the mas to the arcsecond scale. Again, this
suggests the existence of a continuous flow of relativistic electrons
from the core to the arcsecond scale jet. 
The quasar 2134-004 exhibits mas morphology consistent with a core-jet source
suggesting transport of radio plasma to larger scales. However, in this
source there is no evidence on intermediate scales for a continuous
connection. 

\subsection{The nature of GPS quasars}

The relationship between GPS galaxies and quasars has been a
longstanding question. Are the GPS quasars the beamed counterparts 
of the GPS radio galaxies as predicted in Unified Scheme Models
(e.g. Urry \& Padovani 1995)? Or are they unrelated objects
simply having similar radio spectra? 
The idea that the GPS quasars and radio galaxies are different
phenomena has been discussed by Snellen (1997) and Stanghellini et al
(2001).  
Based on the radio morphological evidence presented here on both the
small and large scales, we suggest that the extended emission we
have detected in the GPS quasars is  
currently being supplied with energy by the nucleus. 
A plausible hypothesis is that the GPS quasars exhibiting a core-jet
or complex morphology on the mas scale are indeed ``extended'' (namely
with lobes larger than tens/hundreds of kpc) but core-dominated radio
sources whose redshifts are sufficiently high that most (if not all)
of the the large scale structure is below our detection threshold. 

This would imply that GPS quasars  with core-jet mas morphology
represent  an intermediate population of objects morphologically
more similar to the common flat spectrum radio sources which
dominate the radio surveys at cm wavelengths than to the CSO/galaxy
phenomenon.  If we assume that their jets are oriented with an angle to the
  line of sight intermediate between flat spectrum quasars and radio
  galaxies, GPS quasars would play the same role as the steep spectrum
  radio quasars in the extended sources.

More generally, what causes the GPS
spectral shape in these quasars? \\
There is certainly evidence for moderate beaming in the GPS
quasars (e.g., O'Dea 1998). There is evidence that there is larger 
variability in the GPS quasars than in GPS radio galaxies 
(e.g., Stanghellini 1999; Fassnacht and Taylor 2001; Aller et al. 2002).
In particular, 0738+313, 1127-145, and 2134+004 (in which we find
extended emission) have been shown by Aller et al. (2002) to be
variable. Nevertheless, they usually maintain their convex spectrum
during the variability, although the present shape of the radio
spectrum for 1127-145 is flat and would not be classified as a GPS radio
source today. \\
Furthermore, the GPS quasars have higher polarization than the radio
galaxies (generally unpolarized) but are less polarized than flat
spectrum quasars (e.g., Stanghellini 1999; Aller et al. 2002).  
Hence, it seems plausible that the GPS quasars are not quite as
strongly beamed as the ``flat spectrum" quasars, a possible further
indication of an intermediate orientation.

GPS radio sources are almost evenly optically identified with
galaxies and quasars in the complete sample selected by Stanghellini
et al. (1998).  Does this apply to other samples? 
Dallacasa et al. (2000) selected a sample of extreme GPS radio sources
with convex radio spectra peaking at frequencies above a few GHz, and
with flux density limit at 5 GHz of 300 mJy.
They call these radio sources High Frequency Peakers (HFPs, see also
Dallacasa 2003). That sample extends to higher turnover frequencies
the samples of Compact Steep Spectrum (CSS) and GHz Peaked Spectrum
(GPS) radio sources. Dallacasa et al. (2002a) discuss the optical
identification of the objects in the sample and they find that
26\% are galaxies and 74\% are quasars or stellar objects.

Fanti et al. (2001) selected a new sample of low/intermediate
luminosity CSS radio sources from the B3 sample. The fraction of
quasars is very low and  the vast majority of
the identifications are with galaxies or  empty
fields. 
Dallacasa et al. (2002b,c) observed 46 of these low luminosity 
CSS sources with MERLIN, EVN, and
the VLBA. They find that 36 are CSO or MSO, 2 are core-jet, and 8 of
uncertain or irregular morphology.
Thus, the core-jet morphology on the small scale typically found
in flat spectrum quasars/blazars is found in about half of bright GPS
quasars and only in some faint CSSs. Moreover there is a tendency that
the incidence of quasars and/or core-jet morphologies decreases with
decreasing turnover frequency (i.e., increasing size), and
possibly decreasing brightness.

This is consistent with the view that GPS spectrum in quasars and
galaxies originate from intrinsically different emitting regions:
micro lobes/hotspots with a large range in sizes in galaxies, while
in the quasars the emitting region tends to be more compact and closer
to the core. The structure in the GPS quasars may also be simpler
with most of the emission coming from one or a few knots in a 
moderately beamed jet.  These knots may correspond to the location 
of a shock front where the electron are re-accelerated or a turn in 
a helical jet where the bulk flow is locally pointing toward the
observer, enhancing its brightness by moderate relativistic beaming.
This difference with respect to the other flat spectrum radio
sources may be also the cause for the longer variability time scale
of the flux density reported by Aller et al. (2002)

We remark that some GPS quasars may be young/frustrated/recurrent
radio sources seen with the AGN axis aligned toward us. The mas radio
morphology is fundamental to allow us to test this hypothesis.
However, this may be difficult to achieve in practice since unbeamed
lobe emission may be hard to see {\it in these distant sources}
against the beamed jet/knot emission unless very high dynamic
range images are obtained. 
Generally, if we do not see a CSO morphology  in the quasars, a simple
(and likely) explanation is that the quasar is an intrinsically
large, old and active radio source seen along the radio axis. The
extended emission may not be seen in many cases, since it will be
below our detection limit at the redshift of the quasar. On the
  other hand, CSO morphologies in quasars, or at least mini-lobe
  dominated structures may be signatures of an intrinsically small and
  young radio source.
 
Thus, we suggest that GPS samples are contaminated by radio sources
(mostly quasars) that have nothing to do with the
youth/frustrated/recurrent scenarios. A combination of radio spectrum,
CSO mas radio morphology and the lack of  extended structure should be
used to select young quasars.

\section{Summary}

We searched for extended arcsecond scale radio emission in a sample
of 33 GPS radio sources. We find extended emission which we consider 
likely to be associated with the GPS source in 6 objects. Three of these
have CSO morphology and three have core-jet morphology. At this point
0108+388 remains the only strong candidate for a "re-born" GPS
radio source, with the possible addition of 0941--080.

We suggest that  GPS radio sources with a core-jet mas morphology
(almost exclusively quasars) are likely to be
large radio sources seen in projection and moderately beamed towards us.
 Extended emission is possibly not detectable in some
of these because they are at high redshift where the extended emission
is below our detection threshold. 
GPS quasars are more variable in radio flux density and
more polarized than commonly assumed, and may represent an
  intermediate population between the intrinsically small GPS galaxies
  and the flat spectrum radio quasars.
 Quasars with core-jet or complex morphology tend to be more numerous
 in samples of brighter and smaller radio sources, and dominate the bright
 HFP sample, while they are rare in the intermediate luminosity B3-VLA
 CSS sample. 

We conclude  there is  evidence that GPS/CSS quasars with a
core-jet and/or complex  mas morphology and CSOs are unrelated by
unification, but have similar radio spectra. Any inferences regarding
e.g., radio source evolution based on complete  samples of GPS/CSS
radio sources should take into consideration this contamination  from
the quasars and should be limited to CSOs when information on
morphology is  available or limited to galaxies if mas morphologies
are unknown. 

\acknowledgements{Part of this work has been done during visits
of C.S. at the Space Telescope Science Institute, Baltimore, under
the STScI Collaborative Visitor Program.
The VLA is operated by the U.S. National Radio Astronomy Observatory
which is operated by Associated Universities, Inc., under cooperative
agreement with the National Science Foundation.
The Westerbork Synthesis Radio Telescope is operated by the Netherlands
Foundation for Research in Astronomy (NFRA) which is financially supported
by the Netherlands organization for scientific research (NWO) in the Hague.
We have made use of the NASA/IPAC Extragalactic Database, operated by the
Jet Propulsion Laboratory, California Institute of Technology,
under contract with NASA. This research has made use of the United States Naval Observatory (USNO) Radio Reference Frame Image Database (RRFID).}

\clearpage

\begin{figure}
      \includegraphics[width=8.5cm]{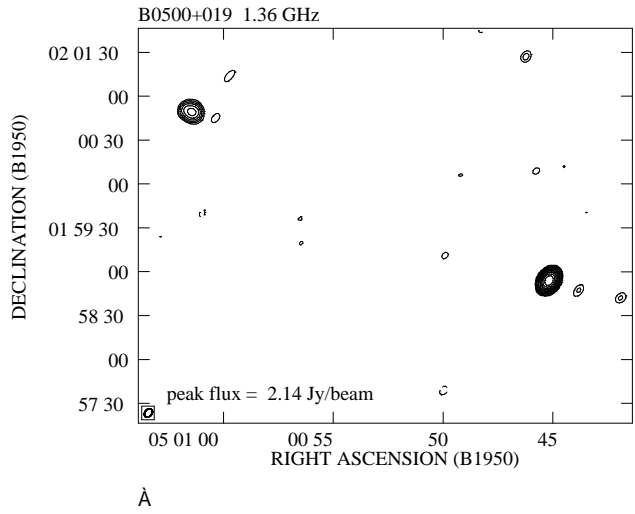}
\caption[]{0500+019 at 1.36 GHz}
\label{0500}
\end{figure}
\begin{figure}
      \includegraphics[width=8.5cm]{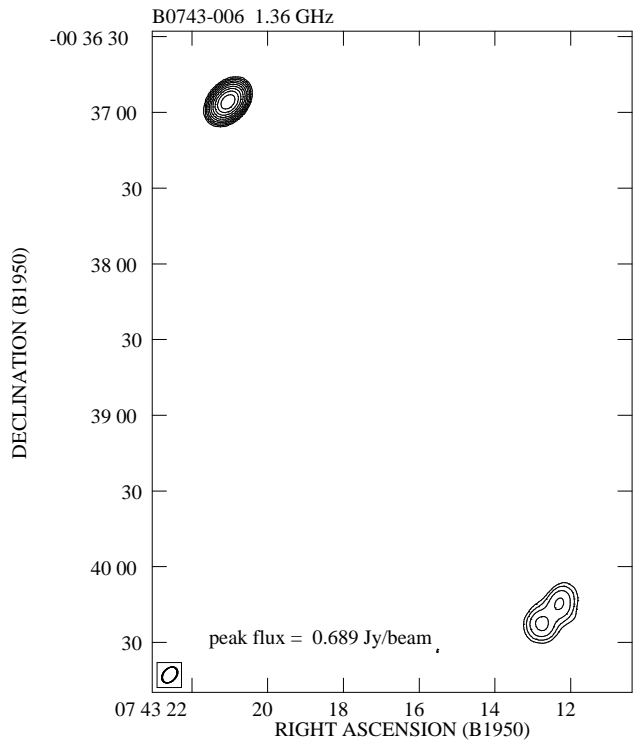}
\caption[]{0743-006 at 1.36 GHz}
\label{0743}
\end{figure}
\begin{figure}
      \includegraphics[width=12cm]{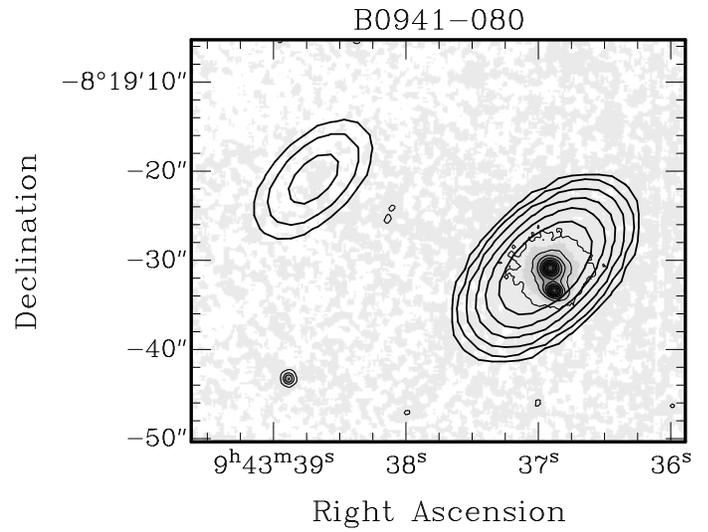}
\caption[]{Radio contours of 0941-080 at 1.36 GHz superimposed to the NOT
optical image (gray scale and thinner contours). First radio contour is 0.5 mJy. 
Peak in the radio image is 2.726 Jy}
\label{0941}
\end{figure}
\begin{figure}
      \includegraphics[width=12cm]{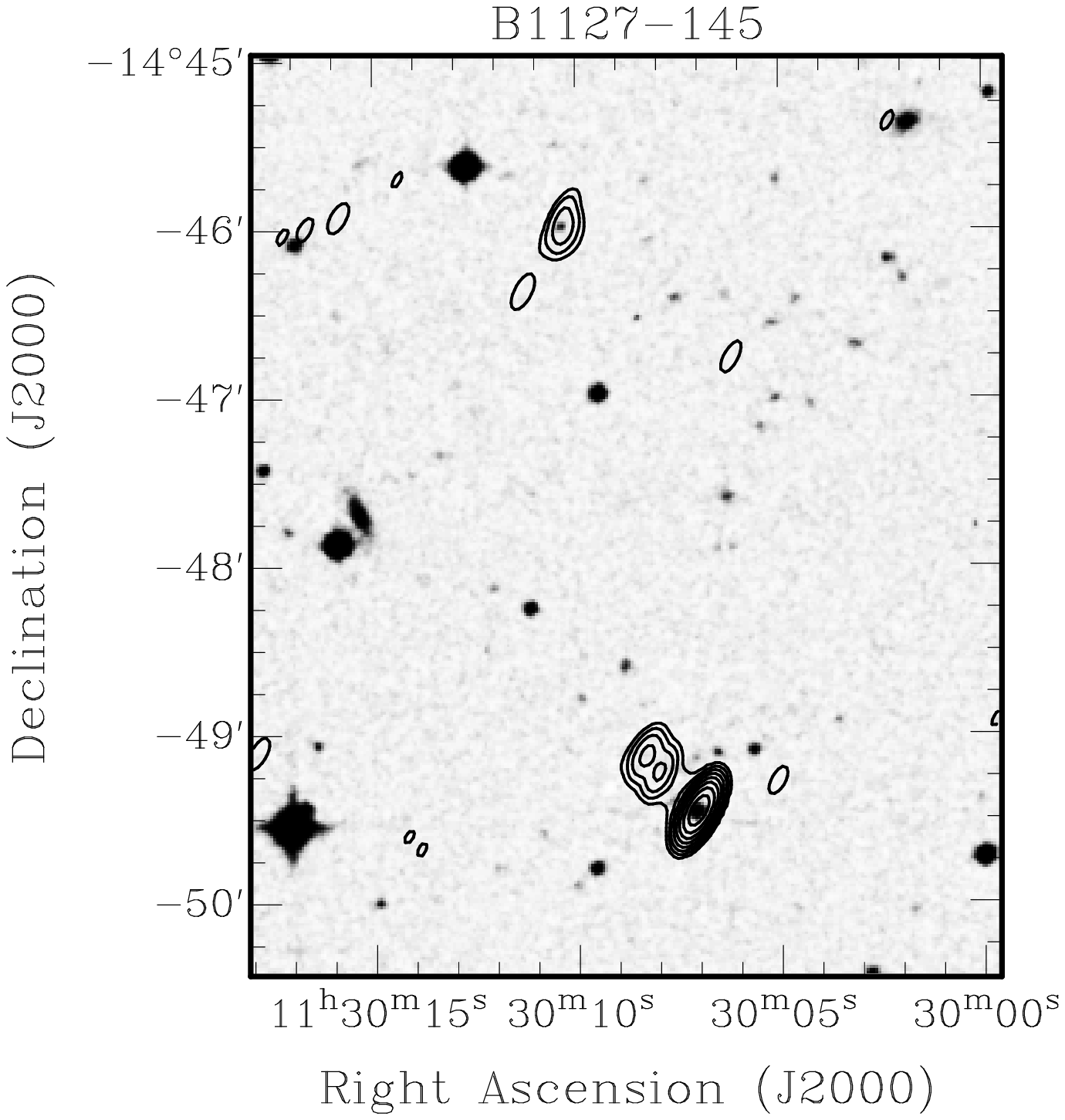}
\caption[]{Radio contours of 1127-145 at 1.36 GHz superimposed to the gray scale POSS2 image}
\label{1127b}
\end{figure}
\begin{figure*}
      \includegraphics[width=18cm]{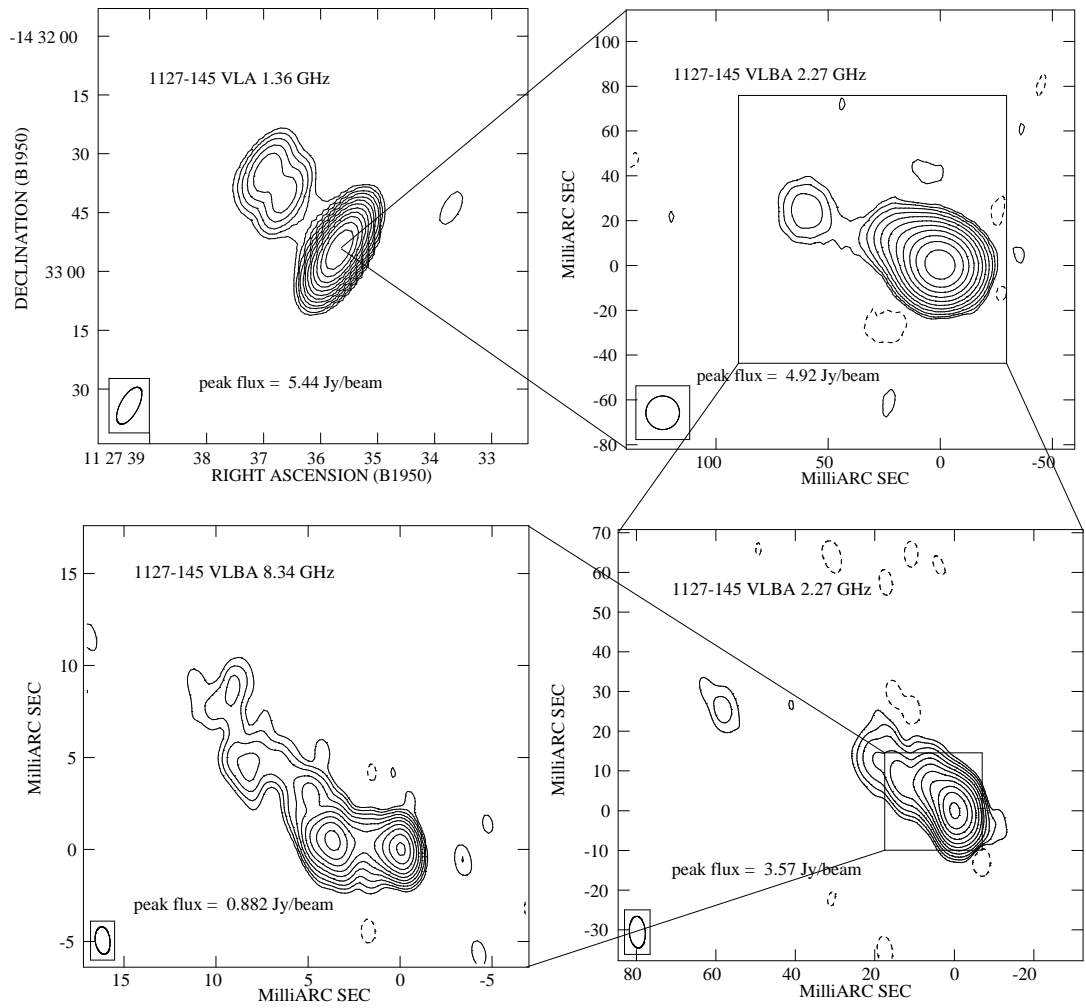}
\caption[]{1127-145 from mas to arcsecond scale}
\label{1127}
\end{figure*}
\clearpage
\begin{figure}
      \includegraphics[width=8.5cm]{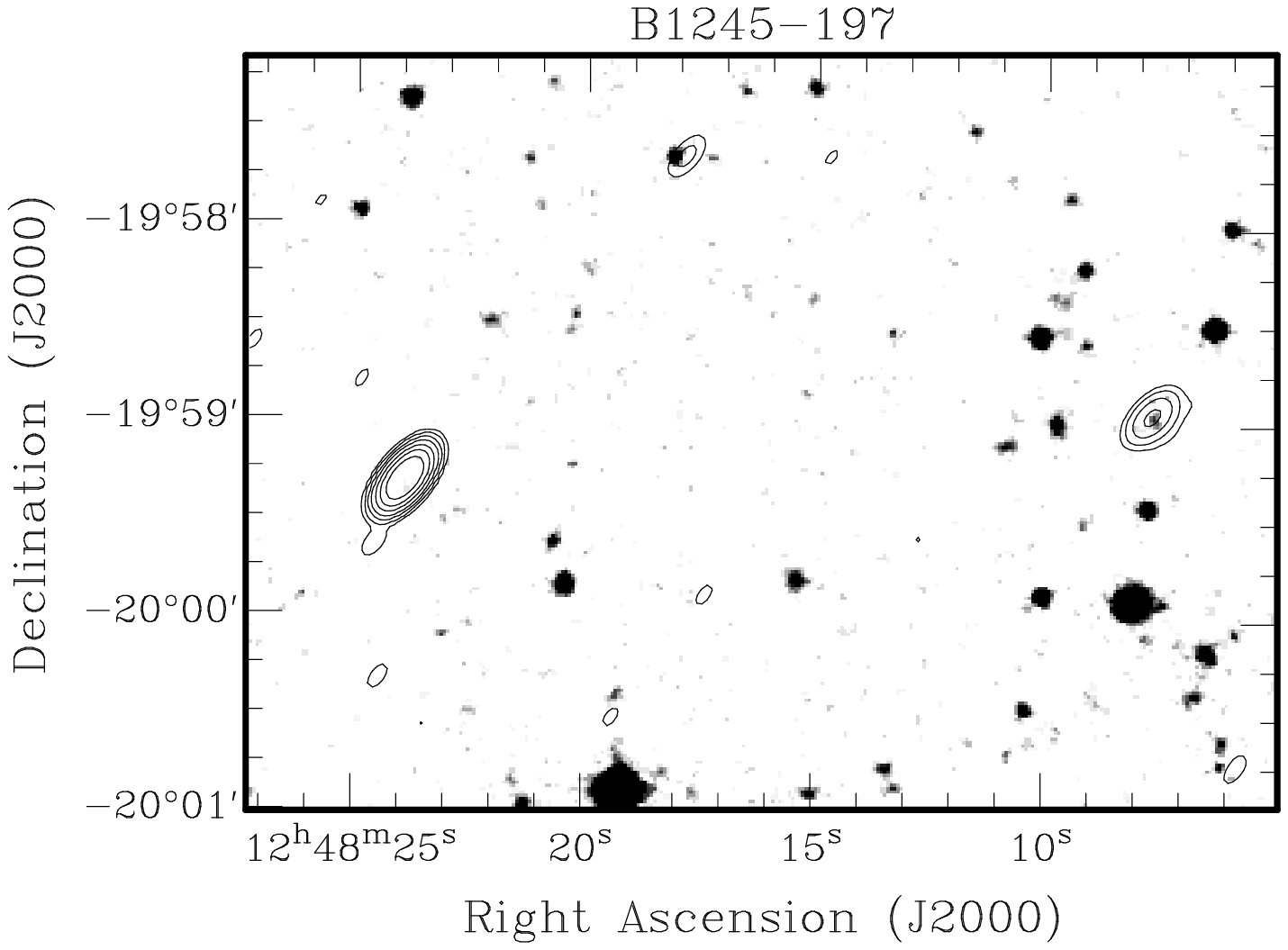}
\caption[]{Radio contours of 1245-197 at 1.36 GHz superimposed to the gray scale POSS2 image}
\label{1245}
\end{figure}
\begin{figure*}
      \includegraphics[width=18cm]{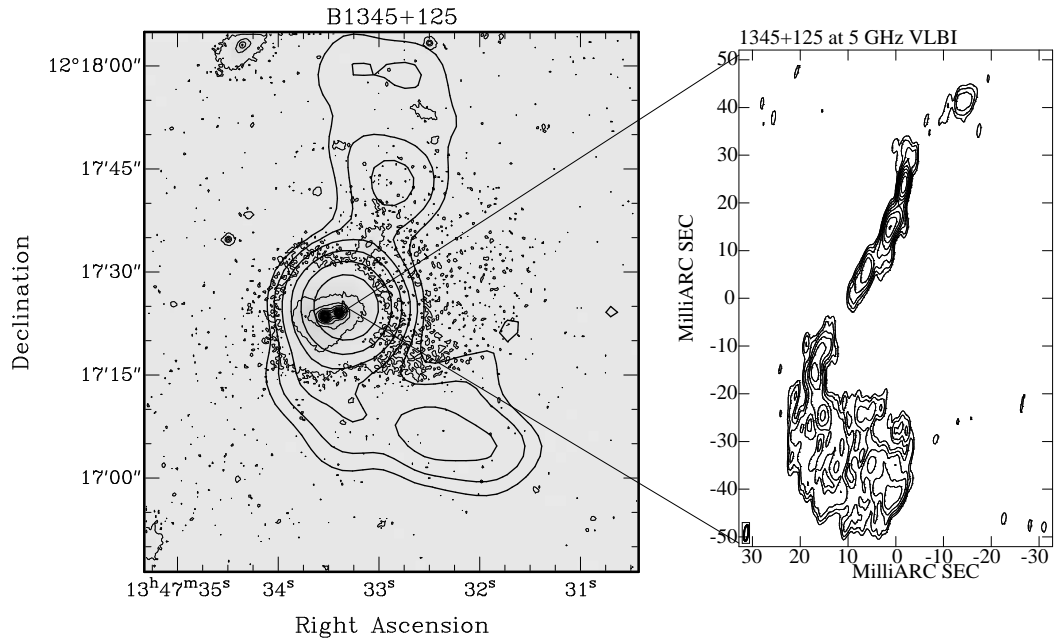}
\caption[]{Radio contours of 1345+125 superimposed to the NOT
optical image (gray scale image and thinner contours). First radio contour is 1 mJy. 
The peak in the image is 64.8 mJy. A point like component
of 5.2 Jy has been subtracted at the peak position to enhance the visibility of 
the diffuse emission around the core.}
\label{1345}
\end{figure*}
\begin{figure}
      \includegraphics[width=9cm]{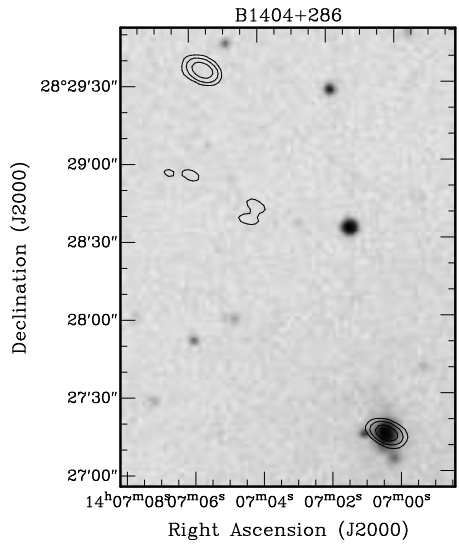}
\caption[]{Radio contours of 1404+286 at 1.36 GHz superimposed to the gray scale POSS2 image}
\label{1404}
\end{figure}
\begin{figure*}
      \includegraphics[width=8.5cm]{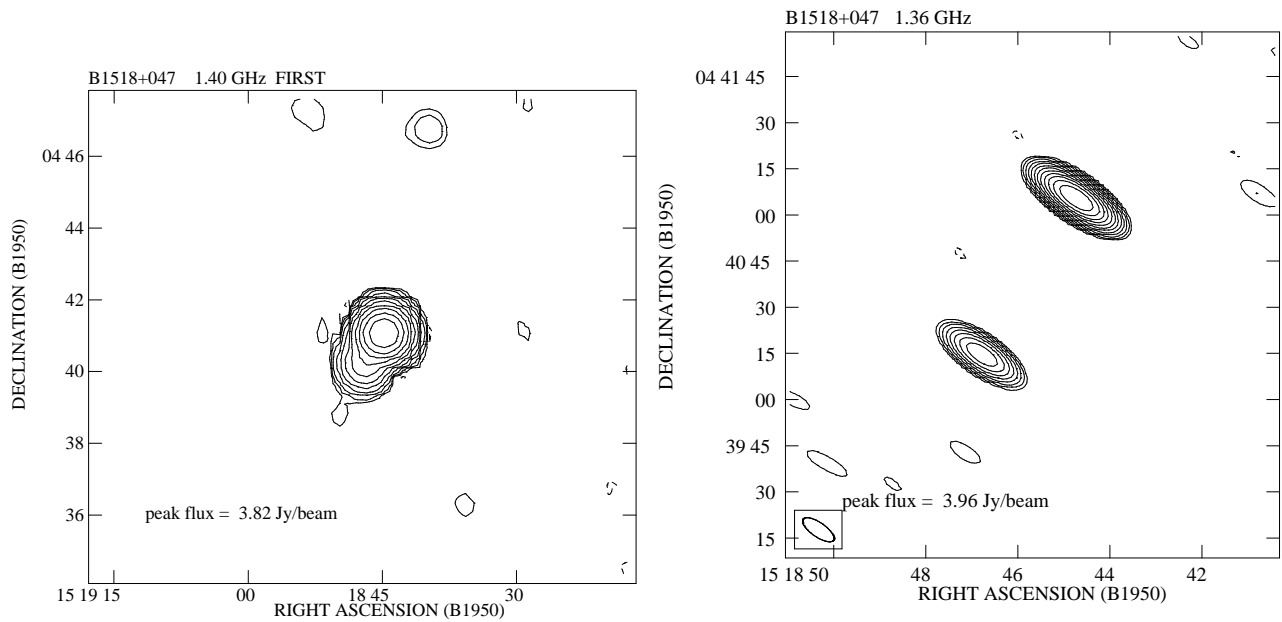}
      \includegraphics[width=8.5cm]{1518_1.4A.PS}
\caption[]{1518+047 from the FIRST (left) and from present data (right)}
\label{1518}
\end{figure*}
\begin{figure*}
      \includegraphics[width=8.5cm]{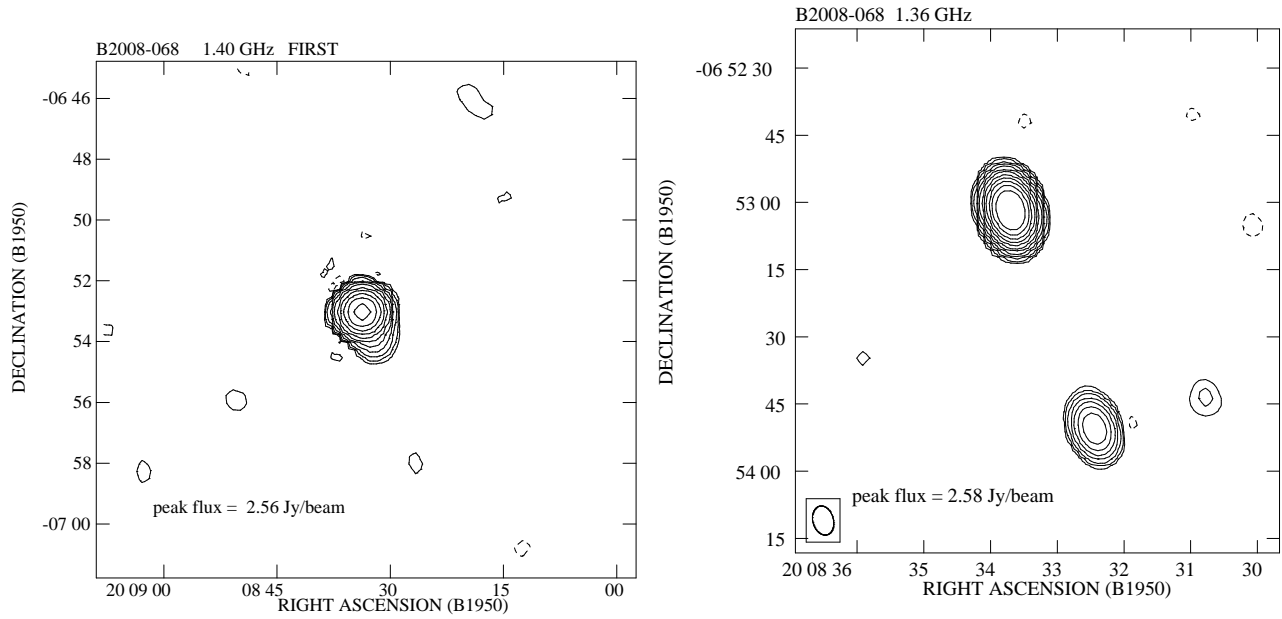}
      \includegraphics[width=8.5cm]{2008_1.4A.PS}
\caption[]{2008-068 from the FIRST (left) and from present data (right)}
\label{2008}
\end{figure*}
\begin{figure}
      \includegraphics[width=8.5cm]{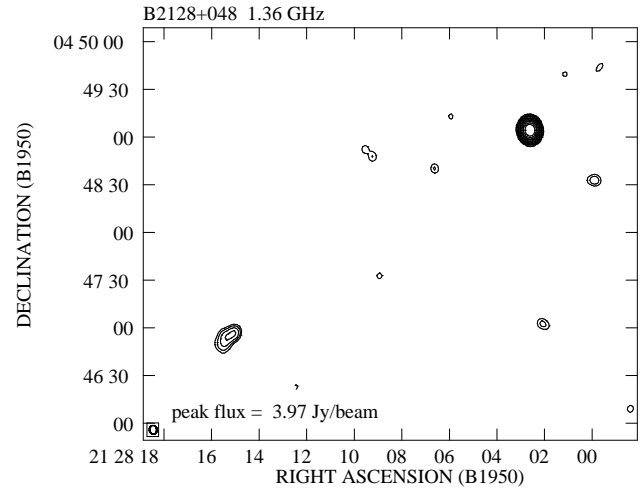}
\caption[]{Radio image of 2128+048 at 1.36}
\label{2128}
\end{figure}
 \begin{figure*}
 \centering
      \includegraphics[width=18cm]{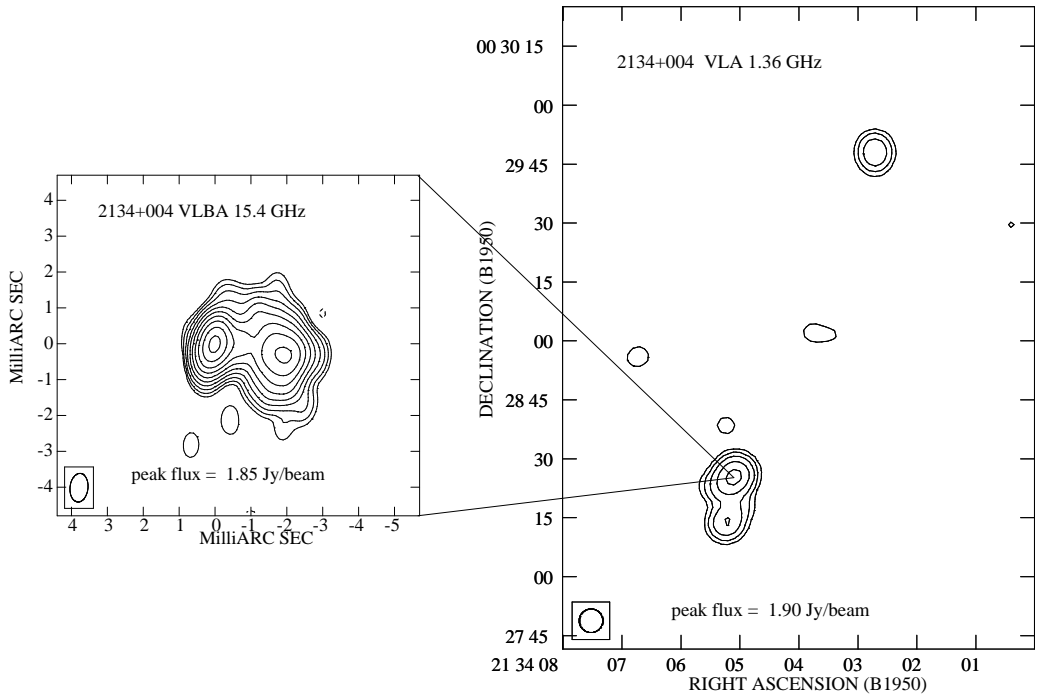}
\caption[]{2134+004 at mas (left) and arcsecond resolution (right). In the VLA 
image most of the core flux density has been subtracted to enhance the visibility of
the diffuse emission.}
\label{2134}
\end{figure*}
\clearpage
\begin{figure*}
      \includegraphics[width=18cm]{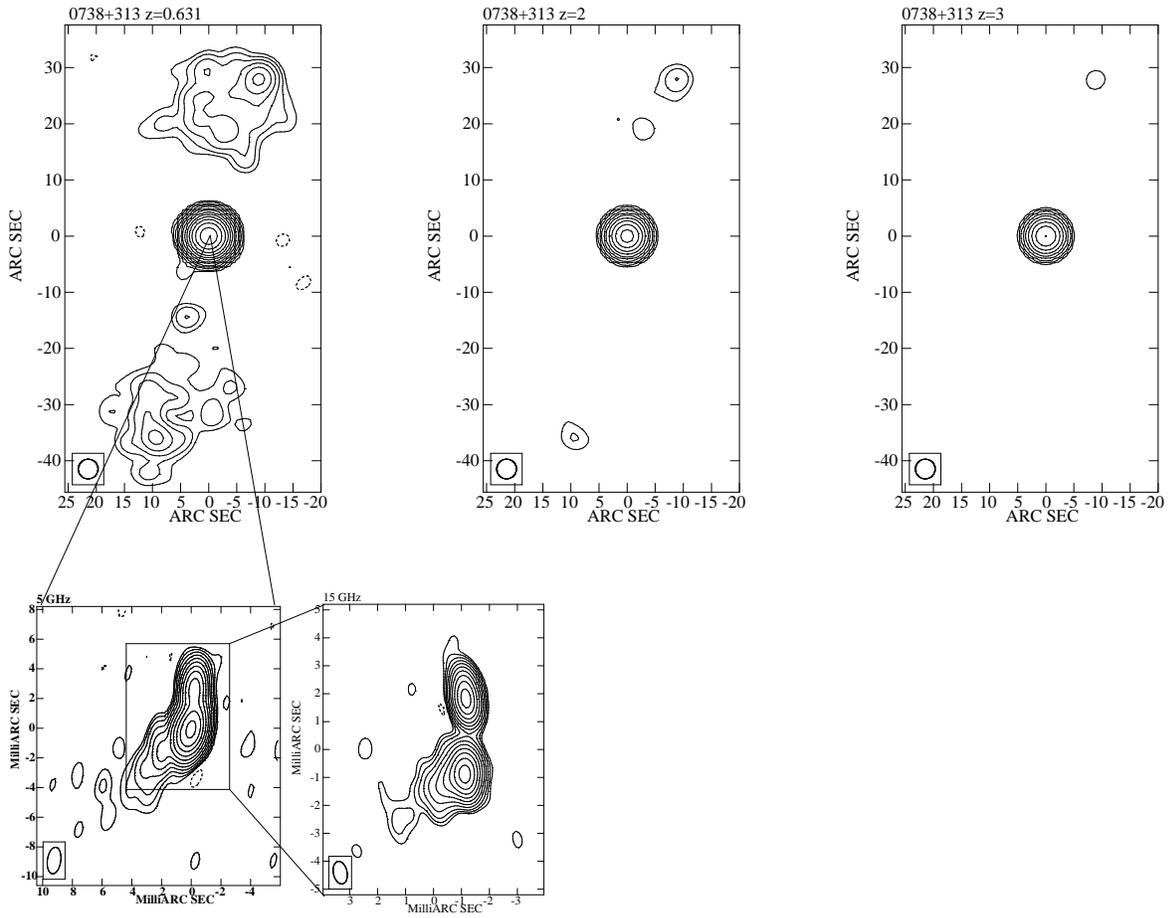}
\caption[]{The effect of redshift on extended emission detectability.
 Left image is 0738+313 at z=0.631, center image is how it would appear at z=2,
 right image is how it would appear at z=3.
 The dependence of angular size with redshift has not been taken into 
consideration because at these high redshifts  angular size does not depend
strongly on  redshift.}
\label{0738}
\end{figure*}

 \begin{figure}
 \centering
      \includegraphics[angle=-90, width=9cm]{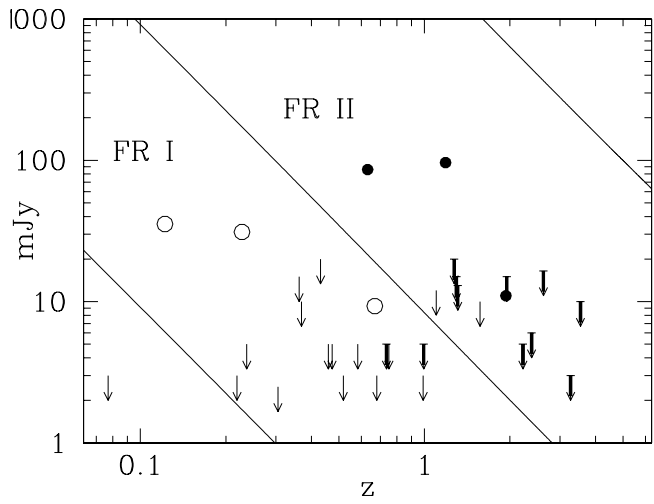}
\caption[]{Flux density at 1.36 GHz of extended emission versus redshift. Continuous lines
approximately separate regions of FRI and FRII radio sources,
filled circles are quasars, empty circle are galaxies, thick arrows
are higher  limits for quasars, thin arrows are higher limits for
galaxies (see section 4).}
\label{fluz}
\end{figure}

\end{document}